\documentclass[twocolumn]{aastex7}

\newcommand{\HeI}{\ion{He}{1}}
\newcommand{\HeII}{\ion{He}{2}}

\begin{document}

\title{A high-resolution $K$-band spectral atlas of massive stars}

\author[]{V. M. Kalari}
\affiliation{Gemini Observatory/NSF’s NOIRLab, Casilla 603, La Serena, Chile}
\email[show]{venu.kalari@noirlab.edu}  

\author[]{W. D. Vacca}
\affiliation{Gemini Observatory/NSF’s NOIRLab, 950 N Cherry Avenue,Tucson, USA}
\email[]{}  

\begin{abstract}
A high-resolution ($\sim$45000), high signal-to-noise ($>$100) $K$-band spectral atlas of massive stars is presented. It includes 81 stars consisting of known optical standards, spanning spectral and luminosity subclasses from O2 to O9, and supergiant luminosity and spectral subclasses from O2--B1. The telluric-corrected reduced spectra are publicly available, and are discussed here. 
\end{abstract}



\section{Introduction} 

Massive, hot stars have a tremendous influence on their environments. Their ionizing radiation and strong stellar winds shape their local environment and enrich the interstellar medium with their nucleosynthetic products. On a global scale, their winds drive outflows from starburst galaxies, and the emission from \ion{H}{2} regions produced by their ionizing radiation serves as a tracer of recent star formation from nearby galaxies up to $z\sim4$ \citep{2014ARA&A..52..415M}. Due to their high effective temperatures (typically$>30000$\,K), their spectral energy distributions peak in the ultraviolet. However, spectra in that regime can only be observed from space-based observatories. Nevertheless, the signatures of winds from massive stars can be observed in ultraviolet spectra of galaxies at moderate redshifts ($z\sim$2-3), which has been shifted into the optical \citep{2022ARA&A..60..455E}.

The stellar initial mass function (IMF) peaks around $0.3\,M_\odot$, with a steep decline at higher masses, meaning that stars more massive than even $\sim15\,M_\odot$ —typical of late-O dwarfs (e.g., O9V) —are relatively uncommon compared to their low-mass counterparts. Coupled with their short lifetimes of only a few million years since their high luminosities drive rapid hydrogen burning-- massive stars are both rare at birth and short-lived. Consequently, they are frequently found within their natal star-forming regions, often heavily reddened by dust, making UV-optical spectroscopy challenging. In our Galaxy, they tend to be found at large distances from us (because of their rarity), and the line-of-sight reddening adds to the local extinction. The extinction toward many known Galactic massive stars exceeds a few magnitudes in the optical.

In such environments, near-infrared (NIR) spectroscopy offers a powerful alternative to the optical and ultraviolet, thanks to its significantly lower extinction coefficients. With the advancement of NIR instrumentation, efforts have been made over the past three decades to spectroscopically classify massive stars in this regime \citep[e.g.,][]{Hanson2005, repolust, 2025ApJS..281...18G, 1996ApJS..107..281H}. These studies have relied largely on low-to medium-resolution ($R\lesssim10000$; or velocity resolutions of $\sim$30\,km\,s$^{-1}$) spectroscopy, which, while valuable, lacks in some cases the resolution required for precision classification and detailed atmospheric modeling.

Although massive stars exhibit sufficient diagnostic features in the $K$ band, including hydrogen Brackett and Pfund series lines, many diagnostic lines, such as the {\HeI} triplet (e.g., at 2.11\,$\mu$m), and the C\,\textsc{4} and N\,\textsc{3} multiplets are blended with other lines. Moderate to high resolution ($R>10000$, although the exact value depends on rotational line broadening) and high signal to noise is needed to separate for example the {\HeI}~2.1614, 2.1617~$\mu$m transitions from the wings of Br$\gamma$ in late type O supergiants and dwarfs, and discern {\HeI} lines in almost all O dwarfs. For e.g., $R\,>$ 25000 is required to separate the {\HeI} 2.112 and 2.1132\,$\mu$m lines for rotational velocities typical of O-type stars (between 50--100\,km\,s$^{-1}$, \citealt{2018A&A...613A..65H}), and this line ratio can be a crucial diagnostic for later O-type stars. 
To the authors' knowledge, there currently exists no collection of massive star spectra in the $K$-band regime at high spectral resolution ($R$ significantly larger than 10000).

The deficiency of a high-resolution NIR spectral atlas of massive stars is especially pertinent given the advent of upcoming 20--30\,m class telescopes that will routinely detect such massive stars in the Milky Way center and in star formation regions in nearby dwarf galaxies \citep{2019arXiv190804687G, 2018arXiv181001738E, 2011BSRSL..80..456E}. A comprehensive spectral reference is essential for interpreting such future observations. To address this need, we present in this paper a high-resolution ($R\sim45000$) NIR $K$-band atlas of known southern massive stars generated with spectra obtained using the Immersion GRating INfrared Spectrometer (IGRINS) at Gemini South \citep[]{igrins1, igrins2}. 

This paper is organized as follows. In Section 2 we present the catalog of sources, and the data reduction procedures. In Section 3, the atlas of sources, and analysis are presented. Section 4 discusses briefly the merits and drawbacks of a NIR spectroscopy with respect to classical optical data, and applicability to more distant sources. The conclusions are presented in Section 5.

\begin{deluxetable*}{lccccccc}
\tabletypesize{\scriptsize}
\tablecaption{Adopted spectral type standards, according to the literature spectral classifications from the aforementioned references. Italics represent stars that conform to those spectral types, but have not been previously used as standards. A detailed list of references for each spectral type, and a discussion of the individual stars is to be found in Section 3. \label{tab:updated_standards}}
\tablewidth{0pt}
\tablehead{
\colhead{} & \colhead{V} & \colhead{IV} & \colhead{III} & \colhead{II} & \colhead{Ib} & \colhead{Iab/I} & \colhead{Ia}
}
\startdata
\hline
O2 & \nodata & \nodata & \nodata & \nodata & \nodata & HD\,93129\,Aa/Ab & \nodata \\
 & \nodata & \nodata & \nodata & \nodata & \nodata & {\it SS\,215/WR20aa} & \nodata \\
\hline
O3 & HD\,64568 & \nodata & {\it ALS\,19312} & \nodata & \nodata & \nodata & \nodata \\
  & {\it THA\,35-II-42/WR21a} & \nodata & \nodata & \nodata & \nodata & \nodata & \nodata \\
\hline
O3 .5& HD\,93128 & \nodata & {\it ALS\,18752} & \nodata & \nodata & HD\,319718 & \nodata \\
 & \nodata & \nodata & \nodata & \nodata & \nodata & {\it ALS\,15210} & \nodata \\
\hline
O4 & HD\,96715 & \nodata & HD\,168076\,AB & \nodata & \nodata & \nodata & \nodata \\
 & \nodata & \nodata & HD\,93250 & \nodata & \nodata & \nodata & \nodata \\
\hline
O4.5 & HD\,303308 & \nodata & \nodata & \nodata & \nodata & {\it ALS\,4067} & \nodata \\
\hline
O5 & HD\,319699 & \nodata & HD\,168112 & \nodata & \nodata & {\it ALS\,1216} & \nodata \\
 & HD\,46150 & \nodata & HD\,93843 & \nodata & \nodata & \nodata & \nodata \\
  & \nodata & \nodata & HD\,93403 & \nodata & \nodata & \nodata & \nodata \\
\hline
O5.5 & HD\,93204 & {\it ALS\,19692} & \nodata & \nodata & \nodata & {\it ALS\,18747} & \nodata \\
\hline
O6 & \nodata & HD\,101190 & \nodata & HD\,152233 & \nodata & \nodata & \nodata \\
 & \nodata & \nodata & \nodata & {\it ALS\,18769} & \nodata & \nodata & \nodata \\
\hline
O6.5 & \nodata & HD\,322417 & HD\,156738 & HD\,157857 & \nodata & \nodata & HD\,163758 \\
 & \nodata & {\it HD\,101298} & HD\,152723 & \nodata & \nodata & \nodata & \nodata \\
  & \nodata & \nodata & HD\,96946 & \nodata & \nodata & \nodata & \nodata \\
\hline
O7 & HD\,93222 & \nodata & \nodata & HD\,151515 & HD\,69464 & HD\,188001/9\,Sge & \nodata \\
& HD\,93146 & \nodata & \nodata & \nodata & \nodata & \nodata & \nodata \\
\hline
O7.5 & HD\,152590/V\,1297\,Sco & HD\,97166 & HD\,163800 & HD\,171589 & HD\,156154 & \nodata & \nodata \\
& HD\,319703A & \nodata & \nodata & \nodata & \nodata & \nodata & \nodata \\
\hline
O8 & HD\,97848 & \nodata & HD\,319702 & HD\,162978/63\,Oph & BD-11\,4586 & \nodata & HD\,151804 \\
\hline
O8.5 & HD\,57236 & HD\,46966 & \nodata & HD\,75211 & {\it HD\,125241} & \nodata & HD\,303492 \\
& HD\,46149 & \nodata & \nodata & \nodata & \nodata & \nodata & \nodata \\
\hline
O9 & \nodata & HD\,93028 & HD\,93249 & \nodata & \nodata & HD\,148546 & \nodata \\
 & \nodata & {\it ALS\,15756} & \nodata & \nodata & \nodata & HD\,152249 & \nodata \\
\hline
O9.5 & HD\,46202 & HD\,155889 & HD\,96264 & \nodata & HD\,76968 & HD\,154368 & \nodata \\
 & \nodata & HD\,96622 & {\it HD\,91651} & \nodata & \nodata & \nodata & \nodata \\
\hline
O9.7 & \nodata & \nodata & HD\,154643 & HD\,152405 & HD \,47432 & HD\,149038/$\mu$\,Nor & HD\,173010 \\
 & \nodata & \nodata & \nodata & HD\,68450  & \nodata & HD\,75222 & HD\,152424 \\
\hline
\hline
B0 & \nodata & \nodata & \nodata & \nodata & HD 150898 & \nodata & HD\,156359 \\
 & \nodata & \nodata & \nodata & \nodata & ALS\,3689 & \nodata & \nodata \\
\hline
B0.5 & \nodata & \nodata & \nodata & \nodata & \nodata & \nodata & HD\,77581/GP\,Vel \\
 & \nodata & \nodata & \nodata & \nodata & \nodata & \nodata & HD\,115842 \\
\hline
B1 & \nodata & \nodata & \nodata & \nodata & \nodata & HD\,91316/$\rho$Leo & \nodata \\
\hline
\enddata
\tablecomments{Except for HD\,152233 and HD\,319718 (from \citealt{martinso}), all O star standards are adopted from \cite{goss1}.}
\label{tab:standards}
\end{deluxetable*}

\section{Observations and data reduction}

\subsection{Catalog of observed stars}
Our targets were selected from O-type standards available in the literature, classified based on optical spectroscopy. They are listed in Table\,\ref{tab:standards}. For O type stars, the classifications in the literature were taken from \cite{2002AJ....123.2754W, goss1, martinso}. For early B supergiants, we adopted spectral types from \cite{2018A&A...616A.149M}. For the selection of standards, we made no distinction between those classified based on the MK schema, line widths (e.g. as in \citealt{goss1}), or via quantitative analysis such as in \cite{martinso}. In total, 81 targets were observed spanning the O2-B1 spectral types, and dwarf to supergiant luminosity classifications. 

Our targets were chosen based on their optical spectral classifications, with the requirements that the final set of sources span the complete set of O2-B1 subclasses and luminosity classes and be visible from the southern hemisphere. The implications of choosing optical spectral standards are discussed in Section 4. Since we were limited to only southern targets, we filled the gaps in the spectral types with optically classified stars from at least medium-resolution spectra for some subclasses. These stars are not usually used as optical spectral standards. They are marked in italics in Table\,\ref{tab:standards}, and the references for their spectral types are given in Table\,\ref{tab:extrastandards}, with a discussion of their spectral type and suitability as standards in Appendix\,A. 

\begin{deluxetable}{lll}
\tablenum{2}
\tablecaption{Observed targets which were not spectral standards}
\tablewidth{0pt}
\tablehead{
  \colhead{Identifier} &
  \colhead{SpT} &
  \colhead{Ref.} \\
}\startdata
  NGC\,3603\,BLW\,A3/ALS\,19312 & O2/3III(f*) & 1\\
  SS\,215/WR\,20aa & O2If*/WN5 & 2\\
      THA\,35-II-42/WR21a & WN5ha/O3+O3Vz((f*)) & 4\\
  ALS\,18752/Cl\,Pismis\,24-17 & O3.5III(f*) & 2\\
  ALS\,15210/Cl\,Trumpler\,16\,244 & O3.5If*Nwk & 2\\
  ALS\,4067/CD-38\,11748 & O4.5Ifpe & 2\\
    ALS\,1216/CD-47\,4551 & O5Ifc & 2\\
  ALS\,19692/[N78]\,49 & O5.5IV(f) & 3\\
  ALS\,18747/HM\,1-6 & O5.5Ifc & 2\\
  ALS\,18769/HM\,1-16 & O6II(f) & 3\\
    HD\,101298 & O6.5IV((f)) & 2\\
  HD\,125241 & O8.5Ib(f) & 2\\
  ALS\,15756/CD-41\,11037 & O9IV & 2\\
  HD\,91651 & ON9.5IIIn & 2\\
\enddata
\tablecomments{$^1$\cite{2002AJ....123.2754W}; $^2$\cite{goss2}; $^3$\cite{2016ApJS..224....4M}; $^4$\cite{2016MNRAS.455.1275T}}
\label{tab:extrastandards}
\end{deluxetable}

\subsection{Observations}
All data were observed in ABBA sequences in a variety of weather conditions, ranging from cloudy and poor seeing (for the bright targets, having $K<7$\,mag) to clear, photometric skies for fainter targets or those in crowded regions. The number of sequences varied depending on the brightnesses of the sources but in all cases the total signal to noise in the observed spectra was greater than 100. 

Telluric standards were taken immediately after or before science observations, closely matching in airmass ($<$0.1). However, due to cloud conditions for some observations, even closely matching tellurics are unable to correct effectively the telluric absorption. This is particularly pertinent near the {\HeI}~2.058$\mu$m feature which suffers greatly from sky emission (see the sky transmission spectrum in Appendix\,B). The observed data also included spectra taken in the $H$-band. These are currently being reduced including further telluric correction, and will be presented in a future publication. The log of our observations is available in Appendix C. 

\subsection{Data reduction}

The data were reduced via the standard IGRINS pipeline v3.2, and are made available via the Gemini archive \footnote{https://archive.gemini.edu/searchform}. The pipeline rejects cosmic rays, corrects for instrumental flexure, and stacks the individual exposures for each slit position (A/B). It then subtracts the stacked A from B nods to remove sky emission features and remove detector readout pattern. Then a flat field (taken on the day of observations except under unusual circumstances) is used to rectify the individual echelle orders. Finally, the pipeline calculates the wavelength solution in vacuum using available sky lines, and extracts the per pixel flux and variance spectra. Telluric absorption is corrected by fitting stellar atmosphere models from \cite{husser} to the observed A0V standard, broadened and shifted to account for rotational and radial velocity of the standard (when known). Additional details are found in \cite{kaplan}. The reader is referred to the pipeline documentation for more details on the data reduction. Note that all wavelengths in this paper are listed in vacuum. 

We found that the final pipeline corrected spectra exhibited significant telluric issues, and were unsuitable for immediate science. We therefore performed further corrections on the delivered datasets. The spectra were corrected for telluric absorption and flux-calibrated using observations of A0V stars obtained at similar air masses and a modified version of the {\textsc{xtellcor}} software \citep{2003PASP..115..389V}. This software assumes  that the observed spectra of the A0V stars can be modeled as scaled versions of the theoretical spectrum of Vega computed by \cite{1992IAUS..149..225K}. The theoretical spectrum is smoothed to the resolution delivered by IGRINS, interpolated to the observed sampling, and then scaled to match the observed magnitudes of the A0V stars. Small shifts (on the order of a fraction of a pixel) were applied to the telluric spectra resulting from the division of the observed A0V spectra by the model in order to account wavelength shifts between the spectra of the A0V stars and the target stars. The latter step greatly reduces residual artifacts in the telluric-corrected spectra of the target stars. The individual orders were then spliced together to generate a single, telluric-corrected, flux-calibrated spectrum of each target star. The spectra were then normalized by fitting a polynomial to the data. 

The raw and pipeline reduced data including the telluric standard spectra is available via the Gemini archive. For some objects, the reduced spectra exhibited unphysical spectral slopes and curvature near the ends of the orders. These are due to the properties of the 
instrument, and make it difficult to normalize the spectra and combine the orders. Hence they lead to periodic jumps in the spectra (see e.g., HD\,152233).

\begin{deluxetable}{lll}
\tablenum{3}
\tablecaption{$K$-band stellar spectral line list adopted}
\tablewidth{0pt}
\tablehead{
  \colhead{Element} &
  \colhead{$\lambda_{\rm Vacuum}$} &
  \colhead{Transition} \\
    \colhead{} &
  \colhead{($\mu$m)} &
  \colhead{} }
\startdata
{\bf{\ion{He}{1}}} & {\bf2.0587} & 2$s{}^1\!S\;\rightarrow\;2p{}^1\!P^0$ \\
\ion{C}{4}$^{1,2}$ & 2.0705 & 3$p{}^2\!P^0_{1/2}\;\rightarrow\;3d{}^2\!D_{3/2}$ \\
\ion{C}{4}$^{1,2}$ & 2.0796 &3$p{}^2\!P^0_{3/2}\;\rightarrow\;3d{}^2\!D_{5/2}$ \\
\ion{C}{4}$^1$ & 2.0842 & 3$p{}^2\!P^0_{3/2}\;\rightarrow\;3d{}^2\!D_{3/2}$ \\
\ion{N}{5}$^3$ & 2.0997 & 1$s^2\,10f\,{}^{2}F^{\circ}_{}
\;\rightarrow\;
1s^2\,11d\,{}^{2}D_{}$ \\
\ion{N}{3}$^1$ & 2.1038 & 8$\rightarrow$7 \\
{\bf{\ion{He}{1}}} & {\bf2.11258} & $4s{}^3\!S\;\rightarrow\;3p{}^3\!P\,$(Triplet)\\
{\bf{\ion{He}{1}}} & {\bf2.11378} & $4s{}^1\!S\;\rightarrow\;3p{}^1\!P\,$(Singlet) \\
\ion{N}{3}$^1$ & 2.1152 & 8$\rightarrow$7 \\
\ion{N}{3}$^1$ & 2.1155 & 8$\rightarrow$7 \\
\ion{N}{3}$^1$ & 2.1156 & 8$\rightarrow$7 \\
{\bf \ion{H}{1} (Br\,$\gamma$)} & {\bf 2.1661} &  {\bf $7 \rightarrow 4$} \\
{\bf{\ion{He}{2}}}$^4$ & {\bf 2.1891} & {\bf $10 \rightarrow 7$} \\
\ion{N}{3} & 2.2471 & $5p\,{}^2P^\circ - 5s\,{}^2S$\,(Doublet) \\
\ion{N}{3} & 2.2513 & $5p\,{}^2P^\circ - 5s\,{}^2S$\,(Doublet) \\
\hline\hline
\enddata
\tablecomments{All wavelengths are retrieved from \cite{NIST_ASD} (NIST database) unless marked. Lines in bold are primary classifiers used here. All wavelengths are quoted in vacuum. $^1$Wavelengths from \cite{1997ApJ...486..420F}. $^2$Note 2.0705 and 2.0796$\mu$m are noted at 2.0802$\mu$m and 2.0842$\mu$m, respectively in \cite{1997PhyS...55..707T}. $^3$Wavelength from  \cite{1997ApJ...486..420F} differs from \cite{NIST_ASD} at 2.0991$\mu$m. $^4$From \cite{1997ESASP.419..273V}. }
\label{tab:linelist}
\end{deluxetable}


\section{Spectral atlas}
\begin{figure*}
\centering
\includegraphics[angle=0,width=1.0\textwidth]{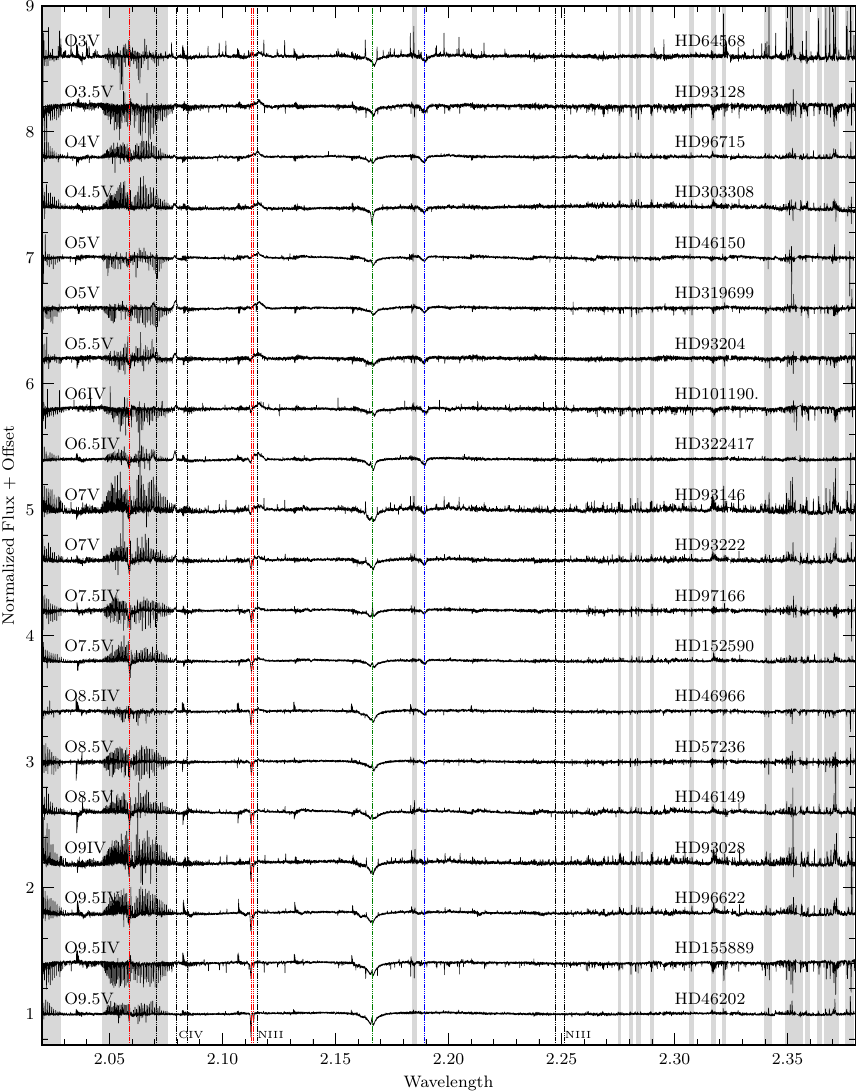}
\caption{Spectral atlas of O dwarf luminosity classes, of known optical standards given in Table\,\ref{tab:standards}. Spectra are normalized, and offset for clarity. Important lines are marked, with red showing {\HeI}, blue {\HeII}, and green for Br$\gamma$ , and metal lines with the element written. All wavelengths are plotted in vacuum, and the complete line list is given in Table\,\ref{tab:linelist}. Two standards, HD\,97848 and HD\,319703 are omitted which is discussed in the text. Some spectra exhibit periodic artifacts (see Section 2.3). The shaded areas marked regions with high sky transmission, $>$0.2 (see also Appendix\,B).}\label{fig:ms}
\end{figure*}

\begin{figure*}
\centering
\includegraphics[angle=0,width=1.0\textwidth]{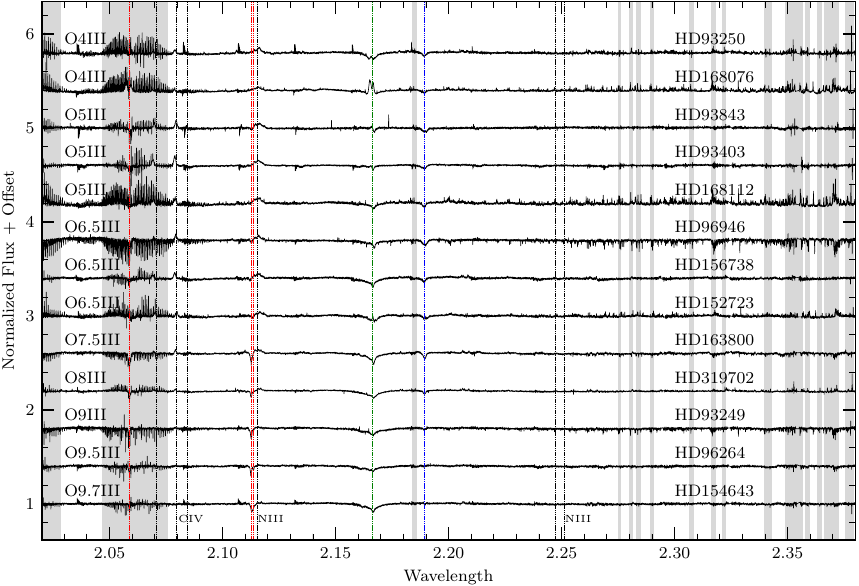}
\caption{Same as Fig.\ref{fig:ms}, but for giants. Note that the spectra of the O3 and O3.5 giants ALS\,19312 and ALS\,18752 are not shown (listed in Table\,\ref{tab:updated_standards}), as these stars are not commonly used optical standards. The spectra are presented and discussed in Appendix\,A. }\label{giants}
\end{figure*}

\begin{figure*}
\centering
\includegraphics[angle=0,width=1.0\textwidth]{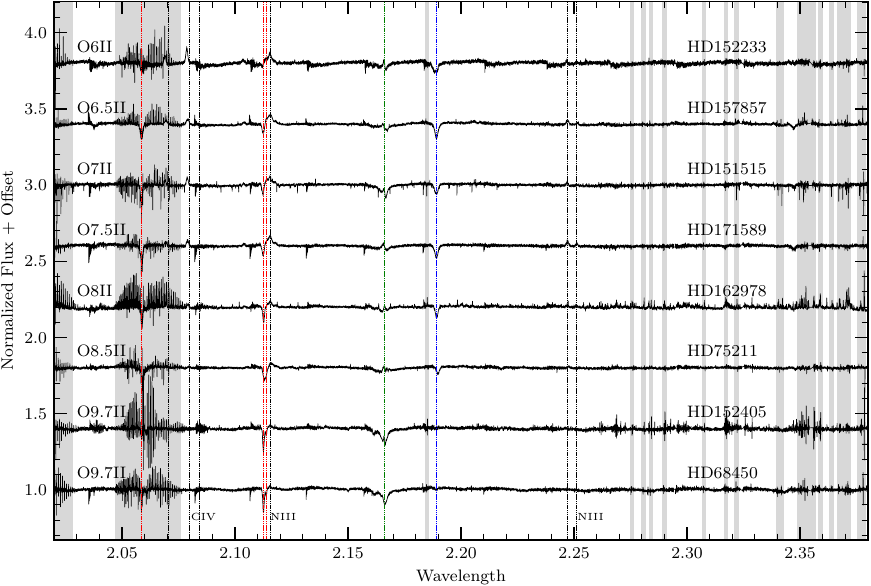}
\caption{Same as Fig.\ref{fig:ms}, but for sub-giants. \ion{N}{3} lines at 2.2471 and 2.2513\,$\mu$m are also shown. }\label{sub}
\end{figure*}

\begin{figure*}
\centering
\includegraphics[angle=0,width=1.0\textwidth]{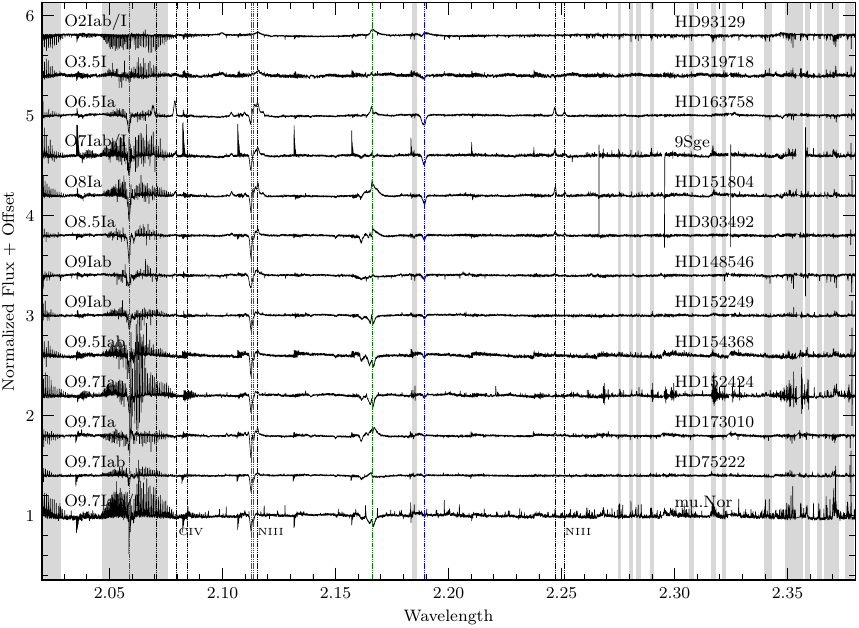}
\caption{Same as Fig.\ref{fig:ms}, but for supergiant Iab/a subclasses. }\label{ias}
\end{figure*}

\begin{figure*}
\centering
\includegraphics[angle=0,width=1.0\textwidth]{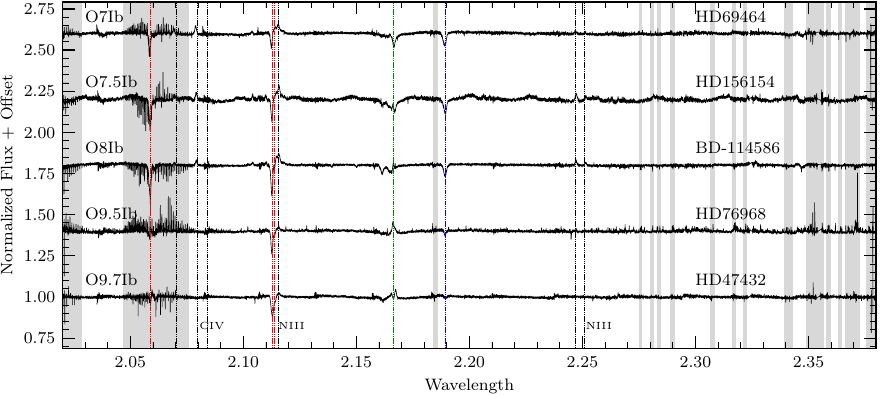}
\caption{Same as Fig.\ref{fig:ms}, but for low luminosity supergiant Ib subclasses. }\label{ib}
\end{figure*}

\begin{figure}
\centering
\includegraphics[angle=0,width=0.5\textwidth]{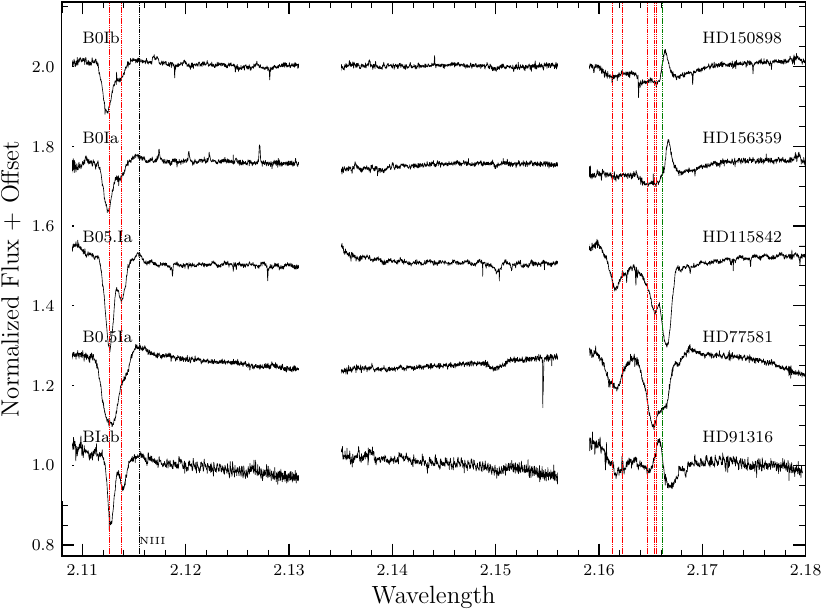}
\caption{Similar to Fig.\ref{fig:ms}, but for early B supergiants. The wavelength is curtailed, and regions with poor corrections are removed, which were particularly strong for some targets, and caused issues during normalization. Additional {\HeI} lines (from Table\,\ref{tab:linelist}) are also marked. }\label{bsgs}
\end{figure}

In Fig.\,\ref{fig:ms}--\ref{bsgs}, we show the $K$-band spectra of all targets given in Table\,\ref{tab:standards}. 
All spectra (telluric corrected) can be publicly downloaded via the zenodo database\footnote{https://zenodo.org/records/19714911}.

In the $K$-band, a few strong lines of {\HeI} and {\HeII} are present, along with Brackett\,$\gamma$ line at 2.16612~$\mu$m. These are listed along with important metal lines in Table\,\ref{tab:linelist}. Also given are the transitions. The wavelengths are listed in vacuum, and are taken either from the National Institute of Standards and Technology (NIST) database for the {\HeI} lines, and from \cite{1997ApJ...486..420F} for the {\HeII} and metal lines. 

\begin{figure}
\centering
\includegraphics[angle=0,width=0.42\textwidth]{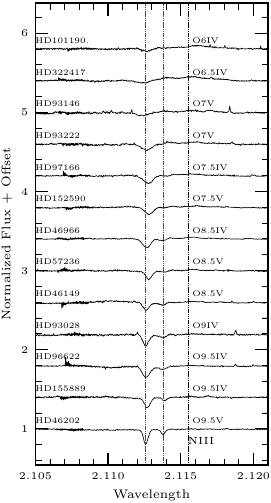}
\caption{Ratio of the {\HeI} singlet (2.113\,$\mu$m) to triplet (2.112\,$\mu$m) which serves as an indicator of spectral type for late type O stars.  }\label{Heiratio}
\end{figure}

\subsection{Spectral classifiers}

\subsubsection{Primary classifiers}
The primary classification lines in the $K$-band for massive stars are the {\HeI}~$2.112~\mu$m triplet and the {\HeII}~$2.1891~\mu$m line (see Figs.\,\ref{fig:ms}-\ref{sgsratio}). High-resolution is particularly necessary to separate the triplet from its nearby singlet at $2.113~\mu$m, as the behavior of these lines is not similar. The singlet is coupled to the UV resonance line at 538\AA\, and can be seen in emission, when the triplet is in absorption \citep{2004A&A...422..275L}. In early to mid-O stars, the nearby \ion{N}{3}~2.115\,$\mu$m lines can 
complicate the estimate of the strengths of the {\HeI} lines, especially in low resolution spectra. In early O dwarfs, {\HeI} is not present, with \ion{N}{3} in emission. The \ion{N}{3} is visible in emission in spectral classes from O3 to O6. In later subclasses the ratio of the {\HeI} singlet to triplet is particularly useful in constraining subclasses, as the singlet strength increases relative to the triplet (Fig.\,\ref{Heiratio}). In contrast, the {\HeII} feature is seen in emission for the earliest spectral types of O2/O3 (for e.g. HD\,93129A, SS\,215, THA\,35-II-42) in our sample, but otherwise appears in absorption till late O spectral types, disappearing by O9. 

Although the Br\,$\gamma$ line varies with spectral type, using this feature for spectral type classification is complicated by contamination by emission from both stellar winds and nebular environments. Unlike Br$\alpha$ which forms close to the photosphere and is an excellent indicator of mass loss, the 7$\rightarrow$4 transition (Br$\gamma$) happens at much lower densities, further out from the photosphere, and therefore is only filled or in emission for the highest mass loss rates. More critical is nebular contamination since the line can be formed in low-density gas, and is frequently found in nebular regions (but with a narrower line profile). In lower resolutions, the numerous {\HeI} lines between 2.158--2.165$\mu$m can also affect the Br\,$\gamma$ line profile. Nebular contamination is present in cases with broad emission wings (the stellar component) and a narrow line profile (the nebular component) as seen in HD\,151804, or narrow emission lines (nebular component) in broad absorption profiles (the stellar component), as observed in HD\,97166, HD\,152590, HD\,57236, HD\,46202). 

Another key line in the NIR is the {\HeI}~$2.058\mu$m line. Unlike other He lines in the NIR, this line descends from the 2$^1 P$ level and so depends strongly on the far-UV 584\AA\ resonance line. In fact, modeling has had difficulties reproducing the observed profiles \cite{repolust}. In addition, this line is located in a spectral region where there is substantial telluric absorption, especially at low observing altitudes. We note that this line is present in most of our spectra (but is not present in the earliest spectral types) with broad absorption profiles. 
However, due to the aforementioned issues we caution against using it as a temperature/luminosity diagnostic as suggested by \cite{2025ApJS..281...18G}. Space observations may better constrain this line strength and its usefulness as a diagnostic of its counterpart UV resonance line strength.  

\subsubsection{Secondary classifiers}

In addition to the strong lines discussed above ({\HeI}~2.058, 2.112, 2.113~$\mu$m, {\HeII}~$2.189~\mu$m, and Br$\gamma$), there exist a plethora of weaker He, and metal lines in the $K$-band spectra of massive stars, detailed in Table\,\ref{tab:linelist}. 

In particular, the \ion{C}{4} triplet around $\lambda$2.07$\mu$m is useful for O stars. The line begins to appear in emission around O4 (but is seen in weak absorption in the earlier types), and disappears around O9, with the 2.0705/2.0796$\mu$m ratio informative, although the former line lies within a wavelength region characterized by poor sky transmission. For mid--late O supergiants, the \ion{N}{3} lines at 2.2471 and 2.2513$\mu$m are usually in emission and can be informative for spectral subclass, with the lines appearing around mid-O and disappearing around late O. This set of lines does not appear in the dwarfs, making them a useful indicator of the giant subclass (preferentially in Ia's). 

We cannot clearly identify the \ion{Mg}{2} lines in late O/early B giants at 2.137, 2.143$\mu$m, as was found by \cite{Hanson2005}. However, the {\HeI} lines between 2.158--2.165$\mu$m are prominent in late O/early B supergiants in absorption, and in weak absorption in late O dwarfs (although even at this resolution this line in the broad Br$\gamma$ profile of dwarfs is blended). 

While the primary and secondary classifiers are useful for most luminosity subclasses from O3--O4 onwards, the most massive stars present relatively few absorption features, and only few strong emission lines. For the earliest stars, even {\HeII}~$2.189\mu$m can be in emission. For O2--O3 stars we suggest classification based on observing the former line. The \ion{C}{4} line appears in absorption at O3. The \ion{N}{3}~2.115~$\mu$m triplet is usually seen as a broad emission feature.

\subsection{Line widths}

\begin{figure}
\centering
\includegraphics[width=0.45\textwidth]{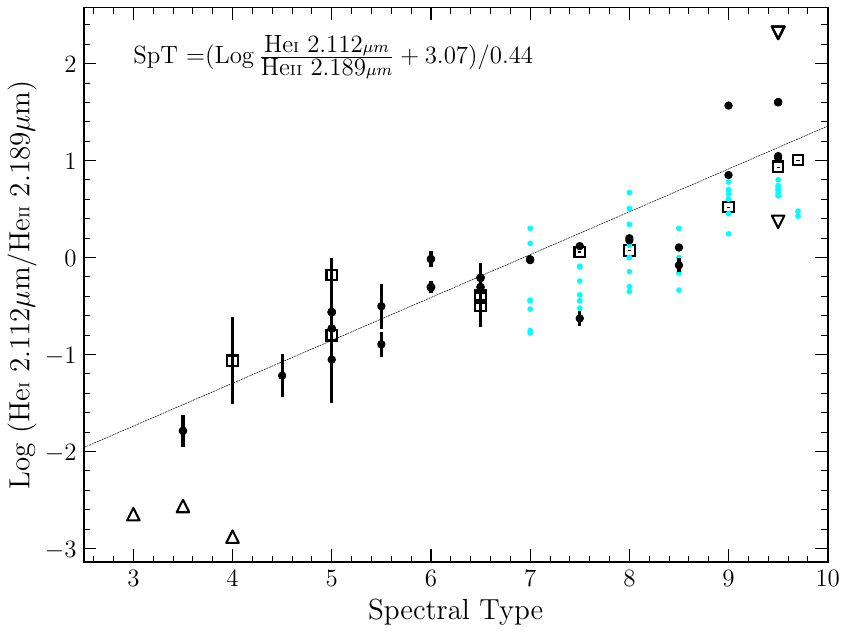}
\caption{{\HeI} 2.1126$\mu$m to {\HeI} 2.1891$\mu$m line ratios versus the optical spectral type are shown for optical standard stars. Dwarfs are shown by circles, while giants are given by open squares. The dashed line is the best least-squares fit to the data. Limits where a given spectral line could not be clearly measured are shown by top (where the {\HeI} is not clearly identified), or down carets (here, the limits are on the {\HeII} line). These were not used in the line-fitting. In cyan are shown the line ratios for all O6--O9 stars from \cite{1996ApJS..107..281H}.}\label{ratios}
\end{figure}

\begin{figure}
\centering
\includegraphics[width=0.45\textwidth]{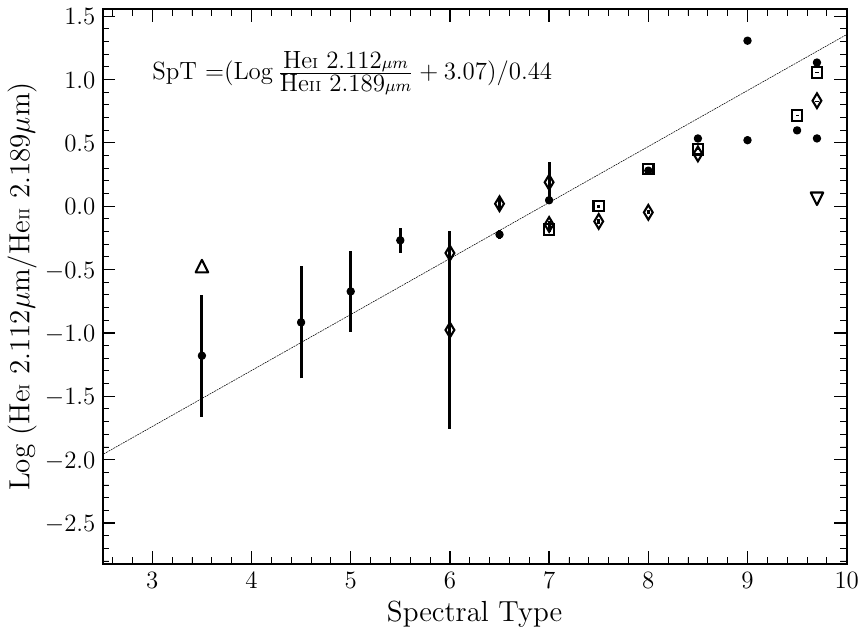}
\caption{Same as Fig.\,\ref{ratios}, but with the Iab/a luminosity classes represented by circles, Ibs by open squares, and sub-giants given by open diamonds. The dashed line is the same as Fig.\,\ref{ratios}.  }\label{sgsratio}
\end{figure}

As discussed in Section\,3.1, we find that the most suitable primary classifier for spectral type of massive stars from NIR $K$-band spectra is the ratio of the {\HeI}~$2.112\mu$m over the {\HeII}~$2.189\mu$m line. The line strength for the former is affected by the weaker triplet in low-resolution spectra (but still is indicative of the photosphere), and in the hottest stars by the rising emission from the nearby \ion{N}{3}~$2.115\mu$m and in rarer cases \ion{N}{5}~$2.1\mu$m emission lines. In contrast, for cooler stars the {\HeII} line is weak or disappears in the later spectral subclasses, and is only clearly identified from mid O7-O8 onwards. 

To estimate the variation of the line ratio by spectral subclass, we measured the equivalent line widths of the two lines for all stars, de-blending where necessary from nearby lines. If {\HeII} or {\HeI} was not clearly present, we determined lower or upper limits on the line ratio, respectively. In Figs.\,\ref{ratios} and \ref{sgsratio} we plot the results for dwarfs and giants, and sub and supergiants respectively. Our line ratios agree well with those found by \cite{1996ApJS..107..281H} for spectral types between O6--O9 (those are the only types in their catalog that have measurements for both lines presented here).

\section{Discussion}

\subsection{Comparison to optical}

Optical spectral standards are well established and have been used successfully to classify objects with unknown spectral types. However, in the $K$-band, the lack of metal lines, the strong He lines, and the telluric absorption and emission present challenges for classification. 

As an example of the difficulties in attempting to carry out hot star classification with only $K$-band spectra, consider the Vz classification introduced by \cite{1973ApJ...179..517W}. This classification is for stars still on the zero-age main sequence, and relies on the presence of the strong {\HeII} 4686\AA~line in the optical spectrum. The lack of optically thin {\HeII} lines in the $K$-band effectively precludes assigning this classification to any hot star based on the $K$-band spectrum alone. Moreover, the 2.058$\mu$m {\HeI}, and the Br$\gamma$ line are highly sensitive to the properties of the stellar wind and are therefore poor discriminators of spectral class.

However, the {\HeI}/{\HeII} 2.112/2.1891~$\mu$m line ratio provides an adequate classification diagnostic, and is similar to the optical {\HeI}/{\HeII} line ratios (for e.g. at 4616/4686\AA\ or 7065/5412\AA\ ). This line ratio is adequate except for the most massive O2--O3 stars where the {\HeII} line may be in emission, and classification must be based on the \ion{N}{5}, \ion{C}{4}, or \ion{N}{3} lines.  In the $K$-band, it becomes essential to have enough resolution to resolve the {\HeI}~2.112$\mu$m from the nearby singlet at 2.113$\mu$m, as these lines can be uncorrelated, and in some cases for the earlier types, they can be contaminated by the strong emission in \ion{N}{3} at 2.115~$\mu$m.

\subsection{Applicability to metal-poor stars}

In general, we expect the spectral standards here to be broadly applicable to more metal-poor stars in nearby dwarf galaxies. They should be particularly useful for young, embedded massive stars found in star-forming regions in the Magellanic Clouds which suffer from high levels of extinction. (for e.g. see \citealt{2014A&A...564L...7K, 2017MNRAS.464.1512W, 2018A&A...615A.121R}).
However, at lower metallicities, the stellar winds are proportionally weaker \citep{2001A&A...369..574V}, which can affect certain lines \citep{2015MNRAS.449.1545B}.

In hot stars, the primary calibrator in the NIR is the {\HeII}~~2.189~$\mu$m line. This line is seen in emission in O2 dwarfs and begins appearing in absorption from O3 onwards. However, it is affected by weakening winds, and we suggest the line should be seen in absorption in earlier spectral types. This expectation is similar to that of the optical lines, as found by recent analysis of statistical sample of optical spectra in the 1/5$Z_{\odot}$ Small Magellanic Cloud by \cite{2023A&A...675A.154V, 2025A&A...695A.198B}, and also by \cite{2014A&A...572A..36T} in local group galaxies, especially given the NIR lines are formed deeper in the photosphere. In such cases, the line ratios shown in Fig.\,\ref{ratios}-\ref{sgsratio} may be inaccurate, and identifications are likely to be tenuous, especially given potentially weaker metal lines. O2--O3 stars in such cases can be classified via the detection 
of \ion{C}{4} triplet (either in absorption or emission) or the presence of  \ion{N}{5}~2.10~$\mu$m.

\section{Summary}

In this paper we present high-resolution high signal-to-noise spectra of 81 massive stars, between the spectral subclasses of O2--O9 spanning dwarf to supergiant luminosity classes, and B0--B1 giant luminosity classes. The telluric corrected spectra are available publicly. 

From this we constructed spectral atlases and concur with previous studies on the NIR spectral diagnostics of massive stars in the literature from qualitative \citep{Hanson2005} and quantitative \citep{repolust} studies. Our main conclusions are- 
\begin{itemize}
    \item There are many lines in the NIR that can be used to study the underlying source that are listed in Table\,\ref{tab:linelist}, but the best stellar photosphere diagnostics are the {\HeI}  2.1125$\mu$m  and {\HeII}~2.1891$\mu$m lines.  
    \item The Br$\gamma$ line is affected by both the stellar wind and underlying photosphere. The {\HeI}~2.058$\mu$m line is a poor diagnostic of the underlying star since it is affected by the wind, and also located in region of poor sky transmission. 
    \item There exist many useful secondary line diagnostics even for the most massive stars, and are applicable to locate such sources in regions of high extinction, or in more distant galaxies. 
\end{itemize}

\begin{acknowledgments}
V.M.K. thanks the Gemini support staff, and the IGRINS team for conducting and reducing these observations, and the anonymous referee for feedback. V.M.K. thanks R. Salinas, H. Kim, A. Guzman for support with obtaining observations, and J. S. Vink for feedback on Section 4.2. V.M.K. and W.D.V. are supported by the international Gemini Observatory, a program of NSF NOIRLab, which is managed by the Association of Universities for Research in Astronomy (AURA) under a cooperative agreement with the U.S. National Science Foundation, on behalf of the Gemini partnership of Argentina, Brazil, Canada, Chile, the Republic of Korea, and the United States of America. This work used the Immersion Grating Infrared Spectrometer (IGRINS) that was developed under a collaboration between the University of Texas at Austin and the Korea Astronomy and Space Science Institute (KASI) with the financial support of the US National Science Foundation 27 under grants AST-1229522 and AST-1702267, of the University of Texas at Austin, and of the Korean GMT Project of KASI.
\end{acknowledgments}

\begin{contribution}

V. M. K. led the observing campaign, analyzed the reduced data, and wrote the manuscript. W. D. V. reduced the pipeline provided data to account for strong telluric features, mis-matching spectral orders, and artifacts; while providing input to the manuscript. 


\end{contribution}

%
\facilities{Gemini (IGRINS)}


\appendix

\section{Notes on selected individual stars}
\noindent

\subsection{Variations in multi-epoch spectra including classifications of non-spectral standard stars}
\begin{figure}[h!]
\centering
\includegraphics[width=1\textwidth]{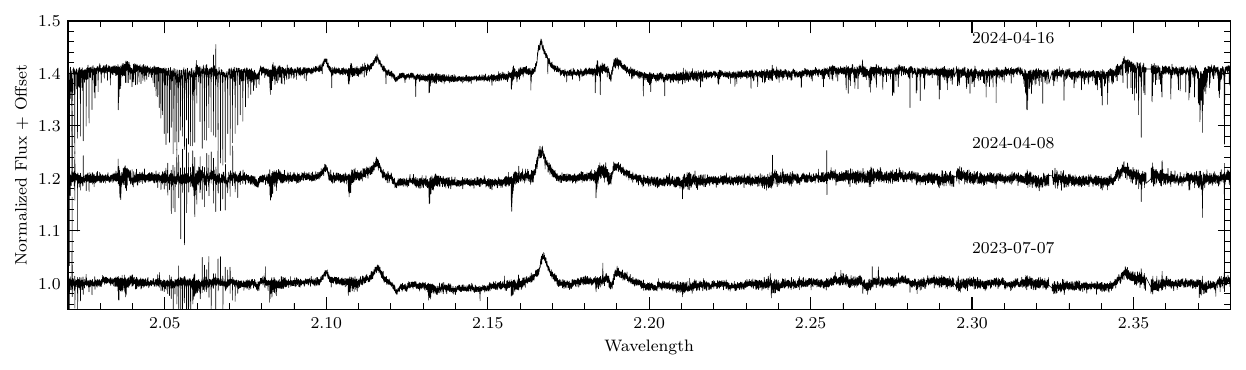}
\includegraphics[width=0.3\textwidth]{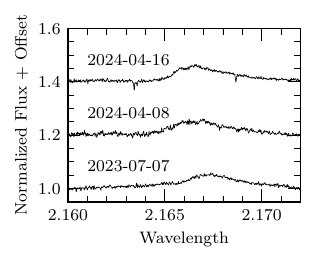}
\caption{{\it Top}: Multi-epoch spectra of HD\,93129A. {\it Bottom}: Br$\gamma$ line profile variations observed in HD\,93129A}\label{93129}
\end{figure}

\noindent
{\bf HD\,93129A} is the archetypal spectral standard Galactic O2I star \citep{2002AJ....123.2754W}. It is actually a binary consisting of Aa and Ab (with spectral types of O2If* and O3IIIf*), and is located in the crowded center of Trumpler 14, and is observed close to periastron epoch (2018.5; \citealt{2017MNRAS.464.3561M}), on 7 July 2023, 8 and 16 April, 2024 (Fig.\,\ref{93129}). The Br$\gamma$ changes remarkably, from a broad red-shifted emission towards a narrower profile (Fig.\,\ref{93129}). Other He and metal lines do not change noticeably. While used as a standard for this spectral type, the change in the Br$\gamma$ line between epochs should be noted, and not considered (as it depends on a variety of factors and not just the underlying photosphere). 

\begin{figure}[h!]
\centering
\includegraphics[width=1\textwidth]{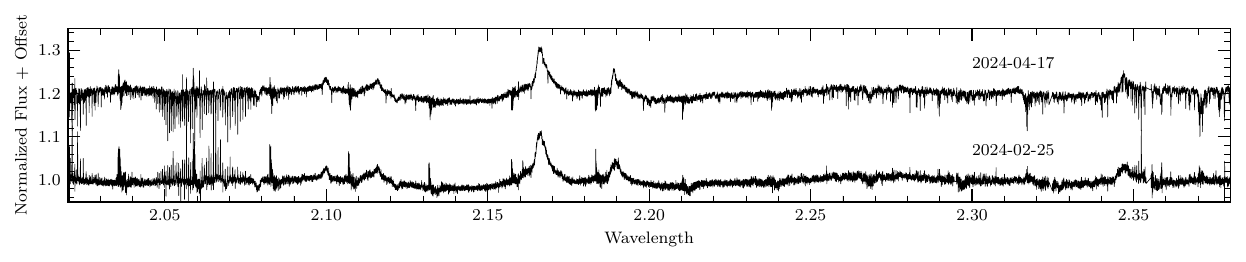}
\caption{Multi-epoch spectra of SS\,215/WR20aa. }\label{ss215}
\end{figure}

\noindent
{\bf SS\,215/WR20aa} was first identified by \cite{2011MNRAS.416..501R}, and is one of the few known O2 (O2If*/WN5) stars in the Galaxy. Two epochs (taken at 12 and 17 April, 2024) show similar profile for the Br$\gamma$ (although small differences exist). The {\HeII} 2.189$\mu$m line is in emission, along with the \ion{C}{4} line in absorption, and the broad \ion{N}{5}~2.10$\mu$m emission (Fig.\,\ref{ss215}). The star appears suited as a reference for slash stars in the NIR given the little change in spectra, and also apparent single nature. 

\begin{figure}[h!]
\centering
\includegraphics[width=1\textwidth]{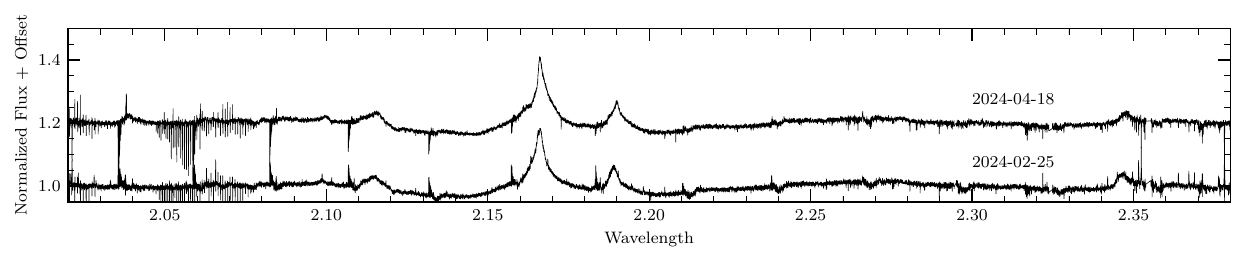}
\includegraphics[width=0.3\textwidth]{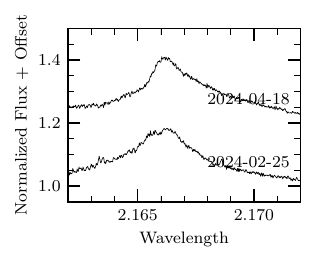}
\includegraphics[width=0.34\textwidth]{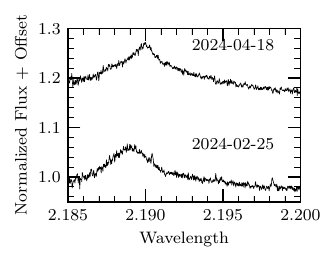}
\caption{{\it Top}:Multi-epoch spectra of WR21a/THA 35-II-42. {\it Bottom left}: Zoom-in of the Br$\gamma$ profile. {\it Bottom right}: Zoom-in of the \ion{He}{2} line profile.   }\label{tha}
\end{figure}

\noindent
{\bf THA 35-II-42/WR21a} is a known O3/WN5ha+O3Vz binary. The orbital period is around 31 days \citep{2016MNRAS.455.1275T}, and the object is unsuited as a spectral standard given its nature, and strong {\HeII} in emission. Two epochs of this star were obtained on 25 February and 18 April, 2024, covering different phases (Fig.\,\ref{tha}). This is seen in the variation of the velocity of He lines, and also the change in Br$\gamma$ line profile.  

\begin{figure}[h!]
\centering
\includegraphics[width=0.32\textwidth]{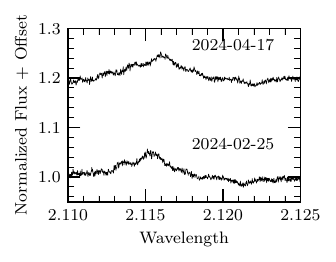}
\includegraphics[width=0.31\textwidth]{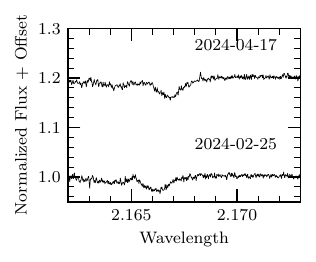}
\includegraphics[width=0.33\textwidth]{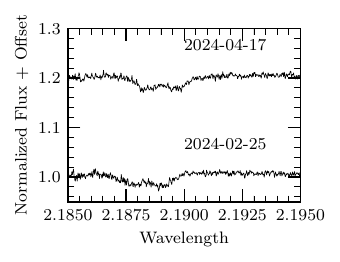}
\caption{{\it Left}: Multi-epoch spectra of ALS\,15210 centered on the \ion{N}{3} triplet, {\it Middle}: Br$\gamma$ profile, and {\it Right}: the {\HeII} line.}\label{tr16244}
\end{figure}

\noindent
{\bf ALS\,15210/Trumpler\,16\,244}
The star is a bright giant, and \cite{goss2} classify it as O3.5If* Nwk star. Our spectra are significantly affected by curvature of the orders. The \ion{N}{3} triplet at 2.1153~$\mu$m are particularly strong in emission, nearly matching the strength of the Balmer emission, agreeing with that classification (Fig.\,\ref{tr16244}). It should be noted that the \ion{C}{4} feature is also prominent around 2.083~$\mu$m. The {\HeII} line at 2.1893~$\mu$m is double lined (the only line visible in absorption), suggesting the potential presence of another star. The second epoch shows varying strength of two line profiles in the {\HeII} feature. Based on these features, the star cannot be considered a representative of it's type in the near-infrared. 

\begin{figure}[h!]
\centering
\includegraphics[width=1\textwidth]{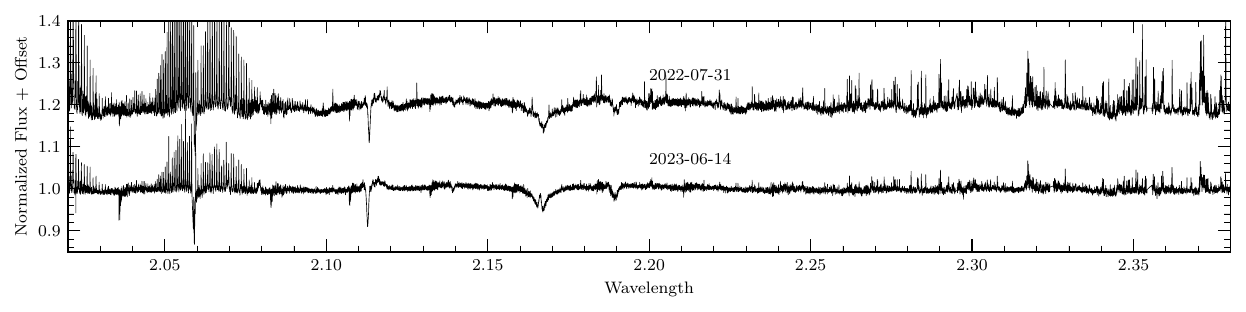}
\caption{Multi-epoch spectra of HD\,152590. }\label{vsconn}
\end{figure}

\noindent
{\bf HD 152590/V1297\,Sco} is a known eclipsing binary \citep{2004IBVS.5495....1O} SB2, which is seen in the $K$-band data. Both epochs are shown in Fig.\,\ref{vsconn} for reference, and the spectral equivalent line widths shown in Fig.\,\ref{ratios} are from the second epoch. 

\subsection{Classifications of remaining non-spectral standard stars and comments}

\noindent
{\bf NGC 3603 BLW A3/ALS 19312} is a very massive star close to the center of the NGC 3603 cluster, and is an O2/O3 IIIf* star as determined by space-based spectroscopy \citep{2002AJ....123.2754W}. Only a single epoch was obtained. Given the slit length of IGRINS (5$\arcsec$), care was taken by the observer to position the slit at 160$^\circ$ such that other bright sources were not in the slit, with the source at the slit end. However, the features and a companion spectrum suggest that the contamination from the nebula and nearby sources was not completely mitigated. The spectrum is given as a reference.

\begin{figure}
\centering
\includegraphics[angle=0,width=1\textwidth]{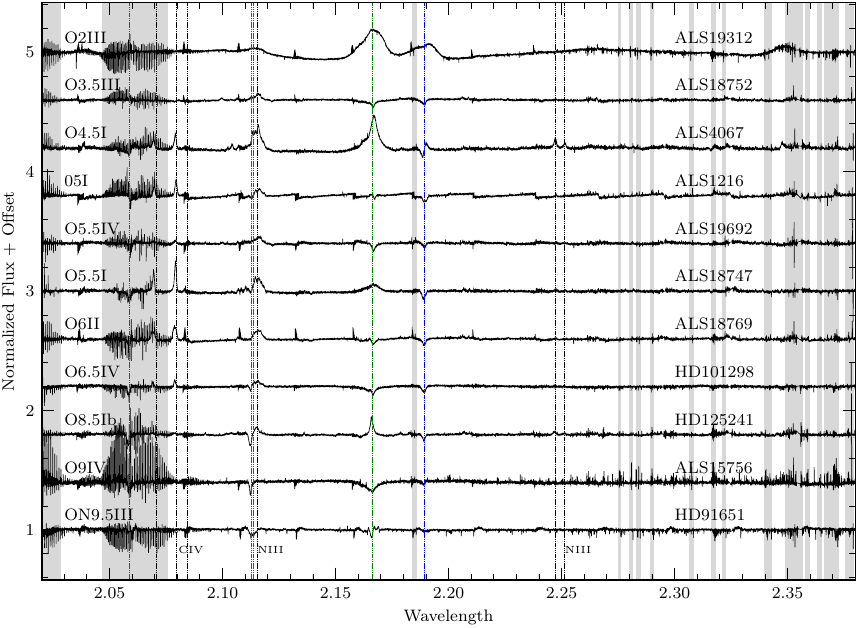}
\caption{Similar to Fig.\ref{fig:ms}, but for the non-spectral standards, along with their literature classification.}\label{leftover}
\end{figure}

\noindent
{\bf ALS\,18752/Cl Pismis\,24\,17} Shows blue shifted absorption in {\HeII}, and strong emission in {\HeI} at 2.112~$\mu$m, along with Br$\gamma$ in absorption. It's spectrum is shown in Fig.\,\ref{leftover}.

\noindent
{\bf ALS\,4067/CD-38\,11748}
Single-epoch spectroscopy of this target suggests it is a suitable spectral standard, as a O4.5I star. It shows an inverse P\,Cygni profile in the {\HeII} line, the \ion{N}{3} lines are visible, and their ratio, along with the strong \ion{N}{3} emission are clear (Fig.\,\ref{leftover}).

\noindent
{\bf ALS\,1216/CD-47\,4551}
The spectrum is affected by normalization issues, and shows strong \HeI emission, with the 2.113~$\mu$m feature in emission. \HeII is in absorption consistent with its spectral type, and the line ratio of \HeI/\HeII is consistent with an O5 or O5.5I classification.

\noindent
{\bf ALS\,19692/[N78] 49}
Marked as an O5.5\,IV, the object displays clear emission in the 2.112~$\mu$m {\HeI} lines, with {\HeII} in absorption. Br$\gamma$ appears in absorption.

\noindent
{\bf ALS\,18747/HM\,1-6}
Shows strong double peaked emission in {\HeI} and Br$\gamma$, but absorption in \HeII.

\noindent
{\bf ALS\,18769/HM\,1-16}
The spectrum is affected by telluric correction issues. While it displays consistent He features with its spectral subclass.

\noindent
{\bf HD\,101298}
The spectral features are consistent with an O6.5\,IV star, and the broad Balmer absorption (affected by nebular emission) is consistent with the luminosity classification. 

\noindent
{\bf HD\,125241}
Is a good example of its spectral class. The Br$\gamma$ emission is strong, with \ion{N}{3} feature still  visible in emission.

\noindent
{\bf ALS\,15756/CD-41 11037}
Another example of its spectral subclass, the \ion{N}{3} is not visible as in other stars of its luminosity classification and subclass, with broad Br$\gamma$ Balmer absorption. The spectra shown here are noisier. 

\noindent
{\bf HD\,91651}
Spectra are again affected by tellurics, with the Br$\gamma$ profile showing inverse P\,Cygni profile. 

\noindent
{\bf HD\,93028} This object was classified as a SB1, with a period around 50 days \citep{1990ApJS...72..323L}, and an O9\,IV by \cite{goss1}. Our $K$-band spectra suggest a later spectral type based on the strength of the {\HeI} lines relative to the {\HeII} lines, but do suffer from blending of the {\HeI} lines at 2.112, 2.113~$\mu$m. 

\noindent
{\bf HD\,97848} Our spectrum is affected by periodic jumps in the spectra. The Br$\gamma$ displays a double line profile, and the \ion{He}{1} line is stronger than expected for it's spectral class, suggesting a later type, and is at odds with it's marginal z classification in \cite{goss2}, where in the \ion{He}{2} would be stronger. We omit it from our final classification figure for this reason. 

\noindent
{\bf HD 155889 AB} Is a known SB2 \citep{2010RMxAC..38...30B}. The broader {\HeII} line affects the measurement of line widths.

\clearpage

\section{Sky transmission spectrum at Cerro Pach{\'o}n}
\begin{figure}[h!]
\centering
\includegraphics[width=1.05\textwidth]{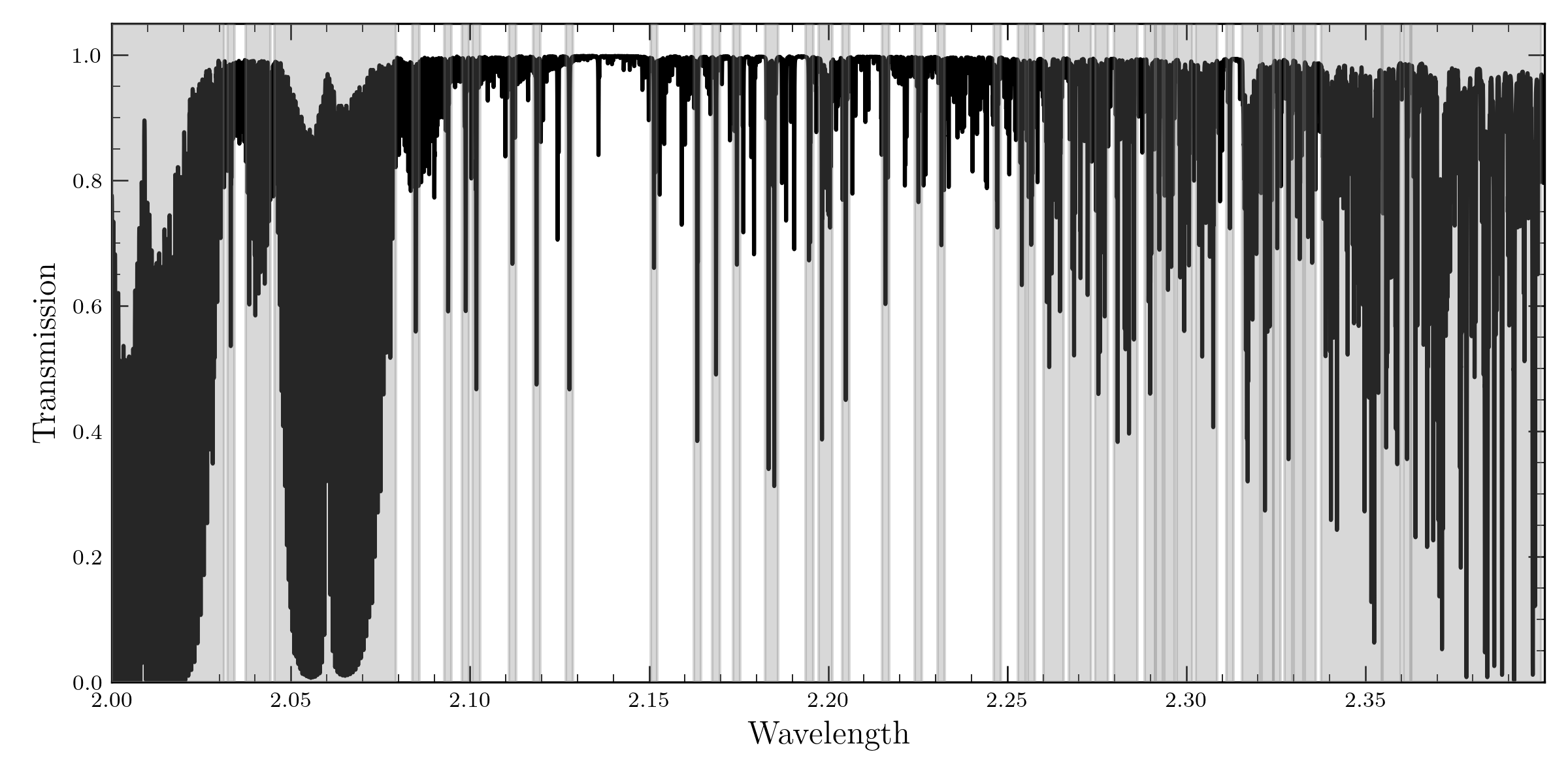}
\caption{Sky spectrum at Cerro Pach{\'o}n. Gray shaded regions are where the sky lines significantly affect the observed spectrum, where the transmission is $<0.4$. Spectral atlases shown have only regions of transmission $>$0.2 marked for clarity. }\label{skytran}
\end{figure}

The sky spectrum as a function of wavelength in the $K$-band is shown in Fig.\,\ref{skytran}. Gray shaded regions mark regions where there are significant sky lines, $<0.4$. It can be clearly seen that lines such as the \ion{He}{1}\,2.058$\mu$m feature, and the nearby \ion{C}{4} triplet are greatly affected by the sky.

\clearpage

\section{Log of observations}

\begin{longtable}{lllllllll}
\caption{Log of observations presented here.}
\\
\hline\hline\
{Date} &
{Name} &
{Spectral Type$^1$} &
{$V^2$} &
{$K$s$^2$} &
{Mean airmass} &
{Exposure Time$^3$} &
{Telluric} &
{Mean airmass$^4$} \\
{} &
{} &
{} &
{(mag)} &
{(mag)} &
{} &
{(s)} &
{ } &
{} \\
\hline
  2024-02-03 & ALS\,1216 & O5I & 8.54 & 5.9 & 1.3 & 10.0 & HIP\,43656 & 1.35\\
  2024-02-25 & ALS\,15210 & O3.5I & 10.71 & 7.06 & 1.19 & 40.0 & HIP\,53836 & 1.2\\
  2023-07-02 & ALS\,15756 & O9IV & 7.9 & 7.4 & 1.32 & 40.0 & HIP\,82560 & 1.28\\
  2024-03-07 & ALS\,18747 & O5.5I & 11.64 & 7.41 & 1.08 & 20.0 & HIP\,86098 & 1.09\\
  2024-02-28 & ALS\,18752 & O3.5III & 11.84 & 6.97 & 1.15 & 60.0 & HIP\,86098 & 1.15\\
  2024-03-01 & ALS\,18769 & O6II & 13.2 & 8.52 & 1.35 & 60.0 & HIP\,86098 & 1.35\\
  2024-04-17 & ALS\,19312 & O3III & 12.95 & 9.31 & 1.18 & 90.0 & HIP55999 & 1.15\\
  2024-03-30 & ALS\,19692 & O5.5IV & 11.87 & 6.56 & 1.0 & 8.0 & HIP\,86098 & 1.01\\
  2024-04-18 & ALS\,3689 & B0Ib & 11.11 & 6.41 & 1.87 & 30.0 & HIP\,81692 & 1.8\\
  2024-03-07 & ALS\,4067 & O4.5I & 11.26 & 6.9 & 1.05 & 25.0 & HIP\,86098 & 1.09\\
  2023-06-14 & BD-11\,4586 & O8Ib & 9.52 & 6.54 & 1.1 & 30.0 & HIP\,93667 & 1.08\\
  2023-07-09 & HD\,101190 & O6IV & 7.33 & 7.16 & 1.86 & 180.0 & HIP\,59851 & 1.78\\
  2024-02-25 & HD\,101298 & O6.5IV & 8.07 & 7.77 & 1.29 & 36.0 & HIP\,55854 & 1.25\\
  2024-04-18 & HD\,115842 & B0.5Ia & 6.09 & 5.18 & 1.48 & 6.0 & HIP\,64933 & 1.41\\
  2023-07-06 & HD\,125241 & O8.5Ib & 8.6 & 6.86 & 1.45 & 60.0 & HIP\,74115 & 1.48\\
  2022-07-31 & HD\,148546 & O9Iab & 7.71 & 6.81 & 1.02 & 24.0 & HIP\,86098 & 1.01\\
  2023-07-06 & HD\,149038 & O9.7Iab & 4.94 & 4.61 & 1.54 & 120.0 & HIP\,81006 & 1.63\\
  2024-04-18 & HD\,150898 & B0Ib & 5.61 & 5.74 & 2.04 & 15.0 & HIP\,82296 & 1.96\\
  2023-06-24 & HD\,151515 & O7II & 7.3 & 6.65 & 1.03 & 16.0 & HIP\,82560 & 1.03\\
  2022-07-31 & HD\,151804 & O8Ia & 5.22 & 4.8 & 1.32 & 5.0 & HIP\,82560 & 1.37\\
  2024-03-09 & HD\,152233 & O6II & 6.59 & 6.1 & 1.1 & 10.0 & HIP\,76618 & 1.08\\
  2023-07-05 & HD\,152249 & O9Iab & 6.45 & 5.75 & 1.15 & 6.0 & HIP\,82560 & 1.18\\
  2023-07-06 & HD\,152405 & O9.7II & 7.29 & 6.8 & 1.08 & 60.0 & HIP\,82560 & 1.11\\
  2023-07-06 & HD\,152424 & O9.7Ia & 6.27 & 5.06 & 1.13 & 40.0 & HIP\,82560 & 1.18\\
  2022-07-31 & HD\,152590 & O7.5V & 9.29 & 7.92 & 1.45 & 60.0 & HIP\,92928 & 1.63\\
  2022-07-29 & HD\,152723 & O6.5III & 7.1 & 6.81 & 1.39 & 30.0 & HIP\,82560 & 1.47\\
  2022-07-29 & HD\,154368 & O9.5Iab & 6.13 & 4.75 & 1.21 & 4.0 & HIP\,83818 & 1.27\\
  2022-07-29 & HD\,154643 & O9.7III & 7.2 & 6.53 & 1.1 & 50.0 & HIP\,86098 & 1.08\\
  2023-06-28 & HD\,155889 & O9.5IV & 6.55 & 6.59 & 1.18 & 17.0 & HIP\,82560 & 1.11\\
  2022-07-29 & HD\,156154 & O7.5Ib & 8.04 & 6.37 & 1.03 & 20.0 & HIP\,86098 & 1.03\\
  2024-04-19 & HD\,156359 & B0Ia & 9.72 & 10.01 & 1.41 & 360.0 & HIP\,80152 & 1.38\\
  2022-07-29 & HD\,156738 & O6.5III & 9.35 & 6.76 & 1.01 & 16.0 & HIP\,86098 & 1.01\\
  2023-07-09 & HD\,157857 & O6.5II & 7.78 & 7.25 & 1.17 & 90.0 & HIP\,94833 & 1.18\\
  2023-07-04 & HD\,162978 & O8II & 6.2 & 5.95 & 1.25 & 15.0 & HIP\,89634 & 1.31\\
  2022-07-31 & HD\,163758 & O6.5I & 7.32 & 7.16 & 1.06 & 60.0 & HIP\,93862 & 1.06\\
  2023-07-07 & HD\,163800 & O7.5III & 7.0 & 6.22 & 1.09 & 100.0 & HIP\,86098 & 1.04\\
  2023-07-03 & HD\,168076AB & O4III & 8.25 & 6.38 & 1.23 & 24.0 & HIP\,88916 & 1.35\\
  2023-07-03 & HD\,168112 & O5III & 8.52 & 6.63 & 1.46 & 30.0 & HIP\,86098 & 1.64\\
  2023-07-04 & HD\,171589 & O7.5II & 8.28 & 7.47 & 1.06 & 50.0 & HIP\,91839 & 1.04\\
  2023-07-04 & HD\,173010 & O9.7Ia & 9.21 & 7.03 & 1.12 & 40.0 & HIP\,93548 & 1.1\\
  2022-08-27 & HD\,188001 & O7Iab & 6.23 & 6.15 & 1.52 & 15.0 & HIP\,95560 & 1.59\\
  2022-03-18 & HD\,303308 & O4.5V & 8.17 & 7.62 & 1.18 & 130.0 & HIP\,49566 & 1.23\\
  2023-07-07 & HD\,303492 & O8.5Ia & 8.85 & 6.95 & 1.28 & 180.0 & HIP\,59898 & 1.28\\
  2023-07-09 & HD\,319699 & O5V & 9.63 & 7.3 & 1.18 & 180.0 & HIP\,88152 & 1.15\\
  2024-02-19 & HD\,319702 & O8III & 10.11 & 7.39 & 1.24 & 40.0 & HIP86098 & 1.26\\
  2022-07-29 & HD\,319703A & O7.5V & 10.92 & 7.03 & 1.01 & 40.0 & HIP\,86098 & 1.01\\
  2022-07-31 & HD\,319718 & O3.5I & 10.44 & 5.89 & 1.05 & 8.0 & HIP\,93862 & 1.05\\
  2023-07-04 & HD\,322417 & O6.5IV & 10.22 & 7.16 & 1.02 & 40.0 & HIP\,82560 & 1.02\\
  2024-02-05 & HD\,46149 & O8.5V & 7.61 & 7.25 & 1.29 & 30.0 & HIP\,28296 & 1.37\\
  2024-02-05 & HD\,46150 & O5V & 6.73 & 6.44 & 1.39 & 15.0 & HIP\,28296 & 1.37\\
  2024-02-19 & HD\,46202 & O9.5V & 8.27 & 7.72 & 1.3 & 62.0 & HIP33297 & 1.34\\
  2024-04-19 & HD\,46966 & O8.5IV & 6.87 & 7.02 & 1.41 & 30.0 & HIP\,33297 & 1.42\\
  2024-02-19 & HD\,47432 & O9.7Ib & 6.31 & 5.86 & 1.22 & 12.0 & HIP30692 & 1.24\\
  2024-02-19 & HD\,57236 & O8.5V & 8.8 & 8.2 & 1.04 & 85.0 & HIP35132 & 1.05\\
  2024-02-04 & HD\,64568 & O3V & 9.39 & 9.12 & 1.01 & 300.0 & HIP\,42334 & 1.05\\
  2022-03-18 & HD\,68450 & O9.7II & 6.44 & 6.47 & 1.03 & 12.0 & HIP\,41576 & 1.04\\
  2022-03-18 & HD\,69464 & O7Ib & 8.8 & 7.77 & 1.07 & 60.0 & HIP\,41576 & 1.1\\
  2023-07-06 & HD\,75211 & O8.5II & 7.5 & 6.4 & 1.65 & 90.0 & HIP\,43571 & 1.66\\
  2022-03-18 & HD\,75222 & O9.7Iab & 7.42 & 6.4 & 1.09 & 40.0 & HIP\,42775 & 1.13\\
  2024-02-04 & HD\,76968 & O9.5Ib & 7.21 & 6.66 & 1.22 & 20.0 & HIP\,42775 & 1.27\\
  2024-04-18 & HD\,77581 & B0.5Ia & 6.87 & 5.6 & 1.03 & 7.0 & HIP\,44395 & 1.03\\
  2024-04-18 & HD\,91316 & B1Iab & 3.87 & 4.28 & 1.3 & 4.0 & HIP\,58510 & 1.25\\
  2024-04-17 & HD\,91651 & O9.5III & 9.52 & 8.81 & 1.24 & 100.0 & HIP\,54231 & 1.16\\
  2023-06-27 & HD\,93028 & O9IV & 8.3 & 8.61 & 1.58 & 160.0 & HIP\,52328 & 1.77\\
  2023-06-27 & HD\,93128 & O3.5V & 8.77 & 7.79 & 1.87 & 62.0 & HIP\,52328 & 1.77\\
  2023-07-07 & HD\,93129A & O2I & 7.88 & 6.01 & 1.43 & 60.0 & HIP\,59851 & 1.43\\
  2023-07-22 & HD\,93146 & O7V & 8.11 & 8.02 & 1.44 & 60.0 & HIP\,53016 & 1.6\\
  2022-03-18 & HD\,93204 & O5.5V & 8.42 & 7.97 & 1.15 & 120.0 & HIP\,59649 & 1.13\\
  2023-07-17 & HD\,93222 & O7V & 8.1 & 7.44 & 1.41 & 40.0 & HIP\,52328 & 1.51\\
  2023-07-09 & HD\,93249 & O9III & 8.2 & 7.99 & 1.8 & 180.0 & HIP\,55332 & 1.75\\
  2024-02-12 & HD\,93250 & O4III & 7.5 & 6.71 & 1.38 & 8.0 & HIP\,53016 & 1.43\\
  2023-07-05 & HD\,93403 & O5III & 8.27 & 6.54 & 1.26 & 60.0 & HIP\,59898 & 1.25\\
  2022-03-18 & HD\,93843 & O5III & 7.33 & 7.23 & 1.25 & 130.0 & HIP\,59898 & 1.24\\
  2023-07-06 & HD\,96264 & O9.5III & 7.62 & 7.79 & 1.52 & 120.0 & HIP\,59898 & 1.29\\
  2023-07-06 & HD\,96622 & O9.5IV & 8.87 & 8.5 & 1.43 & 200.0 & HIP\,53836 & 1.54\\
  2023-07-09 & HD\,96715 & O4V & 8.28 & 7.96 & 1.4 & 90.0 & HIP\,53836 & 1.45\\
  2023-07-09 & HD\,96946 & O6.5III & 8.55 & 7.92 & 1.85 & 180.0 & HIP\,59851 & 1.68\\
  2023-07-17 & HD\,97166 & O7.5IV & 8.91 & 7.69 & 1.43 & 70.0 & HIP\,54830 & 1.43\\
  2023-06-14 & HD\,97848 & O8V & 8.6 & 8.65 & 1.18 & 100.0 & HIP\,48613 & 1.35\\
  2024-02-25 & SS\,215 & O2I & 12.69 & 8.39 & 1.17 & 60.0 & HIP\,53836 & 1.2\\
  2024-02-25 & THA\,35-II-42 & O3V & 12.66 & 7.83 & 1.21 & 30.0 & HIP\,53836 & 1.2\\
  \hline
  \multicolumn{9}{l}{$^{1}$ Adopted in Table 1; 
$^{2}$ Adopted from SIMBAD.
$^{3}$ For a single ABBA sequence
$^{4}$ Mean airmass of telluric standard}
\end{longtable}


\bibliography{ref_subm}{}
\bibliographystyle{aasjournalv7}



\end{document}